\documentclass[12pt,a4paper,notcite,notref]{article}
\usepackage{amsmath}
\usepackage{cite}
\usepackage{graphicx}
\usepackage[usenames,dvipsnames]{color}

\title{On the algebraic structure \\ of rational discrete dynamical systems} 

\author{C-M. Viallet \\  Sorbonne Universit\'es, UPMC Univ Paris 06 \\
 Centre National de la Recherche Scientifique \\
 UMR 7589, LPTHE, 4 Place Jussieu \\ F-75252 Paris CEDEX 05,  France }

\begin{document}
\maketitle

\begin{abstract}
  We show how singularities shape the evolution of rational discrete
  dynamical systems. The stabilisation of the form of the iterates
  suggests a description providing among other things generalised
  Hirota form, exact evaluation of the algebraic entropy as well as
  remarkable polynomial factorisation properties. We illustrate the
  phenomenon explicitly with examples covering a wide range of models.
\end{abstract}

\section{Introduction}

Many of the algebraic aspects of rational discrete systems have
already been investigated, especially in view of their integrability.
A number of these aspects are extensions to the discrete case of
features of continuous systems, but the rationality of the evolution
made the algebro-geometric approach inescapable.  It is indeed at the
basis of any classification attempt, symmetry or multidimensional
consistency analysis, as well as complexity measure via algebraic
entropy. See for example the numerous results exposed in the series of
SIDE meetings~\cite{SIDE}.

The nature of the evolution is also responsible for one recurrent
fact: looking at a finite number of steps of the discrete evolution
yields informations which are in essence of asymptotic nature, like
integrability, hierarchies, or value of the entropy.

Motivated by the original works~\cite{Ko89,Pa02} and comforted by the
more modern approaches~\cite{Ok79,HiKr92} on continuous systems, the
importance of the singularity structure was recognised very
early~\cite{GrRaPa91}.  The use of the apparatus available in two
dimensions, and notably the theory of intersection of curves on
algebraic surfaces, then lead to powerful theorems, in particular on
discrete Painlev\'e equations~\cite{Sa01} and ``QRT''
maps~\cite{QuRoTh88,QuRoTh89,Du10}.

Direct computations of discrete evolutions have also been performed,
especially to detect integrability, endeavouring to reduce the size of
the calculations. For instance, looking at the images of a straight
line in the space of initial conditions, inspired by the geometrical
idea of~\cite{Ar90}, allows to produce an exact sequence of degrees of
the iterates, and in turn to evaluate exactly the algebraic
entropy~\cite{BeVi99,Vi08b}.  Restricting the evolution to integers
lightens even more the calculations. Looking then at the growth of
their height~\cite{Ha05,AnMaVi05}, gives an approximate but efficient
way to evaluate the algebraic entropy. Going even further one may
perform the calculations on finite fields, analyse various statistical
properties~\cite{RoVi03,RoJoVi03}, and eventually detect
integrability.

We take here an opposite attitude, and choose an ingenuous option: we
do evaluate exactly the first steps of the evolutions, and analyse
their structure, and especially their factorisation properties.
\bigskip

The main outcome is that the form of the iterates,
shaped by the singularities, suggests changes of description which
automatically provide:
\begin{itemize}
\item{Exact calculations of the algebraic entropy, and proofs of its
  algebraicity (see~\cite{Si07,Mc07,DiHaKaKo14} for a general point of
  view on this question)}
\item{Generalisations of the discrete Hirota-Sato form and $\tau$
    functions~\cite{Hi77,Sato81,SaSa82,Hi81,Mi82,DaJiMi00,GrRaHi94},
    reducing to the standard quadratic form in specific integrable
    cases}
\item{Various polynomial and integer factorisation properties similar
  the Laurent
  property~\cite{FoZe02,HaPr05,Ho07,FoHo11,Fo14,Prxx}}
\end{itemize}

The plan of the paper is the following:

In section \ref{basics} we briefly recall basic algebro-geometric
notions which will be at the core of the phenomenon we exhibit.

In section \ref{models} we  describe explicitly a number of
models in various dimensions:
\begin{itemize}
\item[ 1 -]{An algebraically integrable two dimensional map in the QRT
    family: McMillan}
\item[ 2 -]{A discrete Painlev\'e equation $dP_{II}$, i.e. a non autonomous
  extension of the previous}
\item[ 3 -]{Another discrete Painlev\'e equation: $qP_{VI}$}
\item[ 4 -]{A  non integrable non confining map in two dimensions: Jaeger}
\item[ 5 -]{A confining non integrable map in two dimensions: JNH-CMV}
\item[ 6 -]{Another  confining non integrable map in two dimensions}
\item[ 7 -]{A three dimensional algebraically integrable map: N=3
  periodic Volterra}
\item[ 8 -]{A linearisable map}
\item[ 9 -]{An unruly model in three dimensions}
\item[ 10 -]{A recurrence of order 2 on functional space, Delay Differential
  equation}
\item[ 11 -]{An integrable lattice map: $Q_V$}
\end{itemize}

We conclude with suggestions for further explorations.

\section{The importance of being singular}
\label{basics}
We use complex projective spaces as spaces of initial conditions
${\cal I}$.  Suppose for simplicity that ${\cal I}$ is of dimension
$N$, with $N+1$ homogeneous coordinates and call $\varphi$ is the
forward map, and $\psi$ the backward map. Then
\begin{eqnarray*}
&& \varphi:[ x_0, x_1,\dots, x_N] \rightarrow [ x'_0, x'_1,\dots, x'_N] \\
&& \psi: [ y_0, y_1,\dots, y_N] \rightarrow [ y'_0, y'_1,\dots, y'_N]
\end{eqnarray*}
The evolution step will always be given by a birational map, so that
both the forward  and the backward evolution are described by
polynomial maps, i.e. $x'_j, j = 0\dots n$ and $y'_j, j=0\dots n$ are
polynomials of degree $d_\varphi$ and $d_\psi$ respectively (usually
$d_\varphi= d_\psi > 1$).

The singular points of $\varphi$ (resp. $\psi$) are the ones for which
$x'_j=0, j=0\dots n$, (resp.  $y'_j=0, j=0\dots n$). We know that the
sets of singular points are algebraic varieties of dimension $\leq
N-2$.

Since `$\psi$ is the inverse of $\varphi$' means that the composition
$\varphi \cdot \psi$ appears as a multiplication of all coordinates by
a common factor, we have the two basic relations
\begin{eqnarray}
\label{kappa}
\psi\cdot \varphi \simeq \kappa_\varphi\cdot id, \qquad \varphi\cdot
\psi \simeq \kappa_\psi\cdot id
\end{eqnarray}
The two polynomials $\kappa_\varphi$ and $\kappa_\psi$, both of degree
$ d_\varphi \, d_\psi - 1$, may be decomposable.
\begin{eqnarray}
\label{kbloc}
 \kappa_\varphi = \prod_{j=1}^{p} (K^+_j)^{l_j}, \qquad \kappa_\psi =
 \prod_{j=1}^{q} (K^-_j)^{m_j}
\end{eqnarray}
 Each factor $K_j^\pm$ defines an algebraic variety of codimension $1$
 playing an important r\^ole in the sequel.

There is a simple relation between the varieties $\kappa_\varphi$ and
$\kappa_\psi$ and the singular locus of $\varphi$ and $\psi$: the
varieties of equation $K^+_j=0$ are blown down by $\varphi$ and their
images are entirely made of singular points of $\psi$. This reflects
the fact that one cannot take one step forward and then a step
backward when starting from a point on $\kappa_\varphi=0$. The same
applies to $\psi$ mutatis mutandis.

Suppose $\Sigma$ is an indecomposable variety of codimension $1$ of
equation $E_\Sigma = 0$. The pullback by $\varphi$ of the equation of
$\Sigma$ gives the equation $E_{\Sigma'}$ of the image $\Sigma'$ of
$\Sigma$ by $\psi$. The important point is that this pullback may
contain additional factors.
\begin{eqnarray*}
\varphi^*( E_\Sigma )= E_{\Sigma'} \, (K_1^+)^{n_1} \, (K_2^+)^{n_2}
\dots (K_p^+)^{n_p}
\end{eqnarray*}
Such factors are necessarily built from components of
$\kappa_\varphi$. Their presence reflects the difference between total
and proper transform: the non-singular points of $\Sigma$ go into the
proper transform. The singular subvarieties contained in $\Sigma$ are
blown up to the other components.  In a way the proper transform is
the true image, disregarding the singularities.

Singularity confinement~\cite{GrRaPa91} in this context is just that
some iterate of $\varphi$ sends components of $\kappa_\varphi$ into
components of $\kappa_\psi$. The regularisation comes, when described
with homogeneous coordinates, from the removal of factor common to all
coordinates. { This is also the origin of the possible drop of degree
  of the iterates of $\varphi$}.

Remark 1: A consequence of the existence of the variety
$\kappa_\varphi=0$ is that the image by $\varphi$ of a generic line
always hits some singular point of $\psi$. The reason is that any
generic line crosses $\kappa_\varphi$, since we are working with
complex projective space.  In other words, while the image of a
generic point is always a generic point, the image of a generic line
is never a generic line.

Remark 2: A variety could be its own transform. It is then covariant
by $\varphi$ and $\psi$. Covariant objects play a fundamental r\^ole
in the description of the algebraic invariants~\cite{FaVi93}.

By abuse of language we will say that $S'$ is the proper transform of
$S$, if $S$ and $S'$ are the equations of varieties which are the
proper transforms of each other by $\varphi$ or $\psi$.

Given an evolution map $\varphi$, denote by $p_k$ the successive
images\footnote{The point $p_{k+1}$ can be obtained from $p_k$ either
  by the action of $\varphi$ on $p_k$, either by pulling back the
  coordinates of $p_k$. The homogeneous coordinates obtained in these
  two ways may differ, but they represent the same point
  projectively.} of $p_0=[ x_0, x_1,\dots, x_N] $. The components of
$p_k$ will factor into indecomposable {\em blocks}. These blocks are
necessarily either the factors $K_i^+$ of the multiplier
$\kappa_\varphi$ and their transforms, either the transforms of the
coordinate planes.  {\em These blocks verify remarkable algebraic
  recurrence relations, and this is the subject of this paper}.

Remark 3: Birational changes of coordinates, which define the natural
equivalence relation between different descriptions of the same model,
affect the singularity structure, the value of $\kappa_\varphi$ and
$\kappa_\psi$, and the form of the equations relating the various
blocks. They however will not spoil the general features of the
recurrences between blocks.

\section{Eleven  models}
\label{models}

The simplest possible type of discrete systems is given by recurrences
of finite order. A recurrence of order $k$ may be looked at as a map
in its $k$ dimensional space of initial data.  There are two natural
generalisations, leading to infinite dimensional space of initial
conditions: recurrences defined over functional space, and recurrences
with multi-indices (lattice maps). Both will be considered, again
supposing rational invertibility of the evolutions.

This section contains the explicit description of the aforementioned
blocks and recurrence relations for eleven different models,
integrable as well as not integrable, finite dimensional as well as
infinite dimensional, to offer a panoramic view on the property we
describe, including  a limiting case (section \ref{unruly}).

\subsection{McMillan}
\label{mcmillan}
The model is a prototype of algebraically integrable map in two
dimension, belonging to the Quispel-Roberts-Thompson
family~\cite{Mc71,Du10,QuRoTh88,QuRoTh89}.

The map $\varphi$ associated to the model reads 
\begin{eqnarray}
\varphi: [ x,y,z] \longrightarrow [-y\, ( {x}^{2}-{z}^{2})+2\,ax{z}^{2},\, x\, ( {x}^{2}-{z}^{2})   ,\, z\, ( {x}^{2}-{z}^{2})  ]
\end{eqnarray}
Its inverse $\psi$ is
\begin{eqnarray}
\psi :  [ x,y,z] \longrightarrow [y \left( y-z \right)  \left( y+z \right) ,x{z}^{2}-{y}^{2}x+2\,ya{z}^{
2},z \left( y-z \right)  \left( y+z \right) ]
\end{eqnarray}
and 
\begin{eqnarray}
\psi \cdot \varphi \simeq \kappa_\varphi =  \left( x-z \right) ^{4} \left( z+x \right) ^{4} = B_1^4\; C_1^4,
\qquad
\varphi \cdot \psi \simeq \kappa_\psi =  \left( y-z \right) ^{4} \left( y+z \right) ^{4}
\end{eqnarray}

The form of the first iterates is\footnote{There always is a rescaling
  possibility of the various factors, in particular an ambiguity
  in the signs, together with a possibility of exchanging $B$ and $C$.}:
\begin{eqnarray}
p_0 & = & [x,y,z] \nonumber \\ p_1 & = & [ A_1, \; x\, B_1\, C_1, \; z
  \,B_1\, C_1] \nonumber \\ p_2 & = & [A_2 \, B_1\, C_1,\; A_1\, B_2\,
  C_2, \; z \, B_1 \, C_1\, B_2\, C_2] \nonumber \\ && \dots \\
 p_k & =& [A_k\, B_{k-1}\,
  C_{k-1},\; A_{k-1}\, B_k\, C_k,\; z\, B_{k-1}\, C_{k-1}\, B_k\, C_k]
\label{stablemcmillan}
\end{eqnarray}

Expressing that $p_{k+1} = \varphi(p_k)$ gives only one condition:
\begin{eqnarray}
( A_k - z B_k C_k)\; (A_k + z B_k C_k)\; ( A_{k-1} B_{k+1} C_{k+1} +
  A_{k+1} B_{k-1} C_{k-1} ) \nonumber \\ - \; 2\; a\; z^2\; A_k \;
  B_{k-1}\; C_{k-1}\; B_k\; C_k\; B_{k+1}\; C_{k+1}
  =0.  \label{mcmillanphik}
\end{eqnarray}
This condition does not suffice to determine  $\{ A_{k+1}, B_{k+1}, C_{k+1} \}$.

{\bf{Claim}}: There exist algebraic relations between the blocks
$A,B,C$.  These relations allow to calculate $\{ A_{k+1}, B_{k+1},
C_{k+1} \}$ in terms of the previous $ A, B, C$'s . Moreover
$A_{k+1}, B_{k+1}, C_{k+1}$ are the proper transforms of $A_{k},
B_{k}, C_{k}$.

{\bf{Proof}}.  We have, for $k\geq 3$:
\begin{eqnarray} \label{hiromcmillan}
\boxed{
\begin{split}
%\begin{cases}
 & A_k - z\, B_k C_k +{\bf B_{k+1}} C_{k-1} = 0, \qquad A_k + z\, B_k
  C_k - B_{k-1} {\bf C_{k+1}} = 0 \\ & A_{k-1}{\bf B_{k+1}} {\bf
    C_{k+1}} + {\bf A_{k+1}} B_{k-1} C_{k-1}+ 2\; a\; z^2 \;A_k\; B_k
  \;C_k =0
%\end{cases}
\end{split}
}
\end{eqnarray}

Equations (\ref{hiromcmillan}) imply (\ref{mcmillanphik}), and can be
verified directly for k=3 and k=4. The validity for general $k$ is
obtained by recursion.   Proving that the form of
(\ref{stablemcmillan}) and of relations (\ref{hiromcmillan}) is stable
is just a matter of counting factors $B$'s and $C$'s. We know from
section (\ref{basics}) that
\begin{eqnarray*}
\begin{cases}
& \varphi^*( A_k) = B_1^{\alpha_B(k)} \;  C_1^{\alpha_C(k)} \; A_{k+1} \\
& \varphi^*( B_k) = B_1^{\beta_B(k)} \;  C_1^{\beta_C(k)} \; B_{k+1} \\
& \varphi^*( C_k) = B_1^{\gamma_B(k)} \;  C_1^{\gamma_C(k)} \; C_{k+1} 
\end{cases}
\end{eqnarray*}
for some exponents $\alpha, \beta, \gamma$.  Using
(\ref{hiromcmillan}) we get $A_{k+1}$ from the previous $A, B,
C$'s. We may thus evaluate all the exponents $\alpha(k+1), \beta(k+1),
\gamma(k+1)$. The outcome is that {\em $A_{k+1}, A_{k+1},B_{k+1}$ are
  the proper transforms of $A_{k}, A_{k},B_{k}$} and do not
factorise, as no new factors $B_1$ or $C_1$ are left over in the
components of $p_k$ after $k=3$. QED.

%{\bf Definition} We will call 

We will present similar properties in the subsequent sections. Their
proof goes along the same lines and will not be detailed.

Relations (\ref{hiromcmillan}) define completely the evolution of $\{
A_k, B_k, C_k\}$.  They extend over a string of points of length $3$.
Although their solution is written as fractions, the result is
automatically {\em a polynomial in terms of the initial conditions $
  [x,y,z]$}.  They moreover enjoy a Laurent
property~\cite{FoZe02,HaPr05,Ho07,FoHo11,Prxx}.

 Define a map $f_\varphi : [ P,Q,R,U,V,W ] \longrightarrow
 [U,V,W,U',V',W']$ with
\begin{eqnarray*}
 U'= \frac{ -P\; V'\; W' -2\; a\;z^2\; U\; V\; W}{Q\, R}, \quad V' =
 -\frac{U-z\, V\, W}{R}, \quad W' = \frac{U+z\, V\, W}{Q}
\end{eqnarray*}
which implements the solution of (\ref{hiromcmillan}) as a map.  We
may consider iterations of $f_\varphi$ starting from arbitrary initial
data $[p,q,r,u,v,w]$. The images are Laurent polynomials in
$[p,q,r,u,v,w]$.  If in addition the triplets $[p,q,r]$ and $[u,v,w]$
happen to be of the form $[ A_{k-1}, B_{k-1}, C_{k-1}]$ and $[ A_{k},
  B_{k}, C_{k}]$, then the iterates are { polynomials}.

Setting $\Gamma_k = B_k C_k$, one may rewrite the 'raw' form 
(\ref{hiromcmillan}) as
\begin{eqnarray*}
\begin{cases}
&  A_k^2 - z^2\, \Gamma_k^2 + \Gamma_{k-1} \Gamma_{k+1} =0 \\
&  A_{k-1} \Gamma_{k+1} + A_{k+1} \Gamma_{k-1} + 2\, a\, z^2 A_k \Gamma_k =0
\end{cases}
\end{eqnarray*}
which is nothing but the bilinear quadratic discrete Hirota form, the
inhomogeneous coordinates of the $k$th iterate being just $
[A_k/(z\Gamma_k), A_{k-1}/(z\Gamma_{k-1})]$, as suggested
in~\cite{RaGrSa95,OhRaGrTa96,KaMaOi00}, following~\cite{Pa02,HiKr92}.

The recurrence (\ref{hiromcmillan}) also gives the constraints obeyed
by the sequence of degrees $\delta(A_n), \dots$ of the successive
$A,B,C$'s,
\begin{eqnarray*}
\begin{cases}
  &\delta( A_{k+1} ) + \delta( B_{k-1} \, C_{k-1}) = \delta ( A_{k-1}
  ) + \delta( B_{k+1}\, C_{k+1} ) = \delta( A_k) +\delta(B_k \,C_k) +2
  \\ & \delta( B_{k+1} ) + \delta( C_{k-1} ) = \delta( C_{k+1}) +
  \delta( B_{k-1}) = \delta ( A_{k} ) = \delta( B_k\, C_k) +1 \\
 \end{cases}
\end{eqnarray*}
and consequently the one verified by the degree $d_n = \delta(A_n) +
\delta(B_{n-1} \;C_{n-1})$ of $p_n$
\begin{eqnarray*}
d_n -3\, d_{n-1} + 3\, d_{n-2} - d_{n-3} = 0,
\end{eqnarray*}
which proves  quadratic growth of $\{ d_n\}$ and vanishing of the
algebraic entropy.

Finally the invariant of the model may be rewritten
\begin{eqnarray*}
  I = 
  \left( \frac{A_k}{z \Gamma_k}\right)^2 + \left(\frac{A_{k+1}}{z \Gamma_{k+1}}\right)^2 
  - \left( \frac{A_k A_{k+1}}{ z ^2 \Gamma_k \Gamma_{k+1}} -a  \right)^2.
\end{eqnarray*}

%%%%%%%%%%%%%%%%%%%%%%%%%%%%%%%%%%%%%%%%%%%%%%%%%%%%%%%%%%%%%%%

\subsection{$dP_{II}$}
This an integrable non autonomous extension of the previous model, and
has the Painlev\'e~${II}$ equation as a continuous
limit~\cite{SaKaGrHiRa95}. Its non autonomous nature invites us to
write the map in three dimensions, the added variable having a linear
evolution.
\begin{equation*}
\varphi: [x,y,z,t) \longrightarrow 
[- y(t^2 -{x}^{2} )+ \left( c\,z+b\, t \right) xt ,
x \left( t^2-x^2 \right)   ,
 \left( z+t \right)  \left( t^2-x^2 \right)  ,
t \left( t^2-x^2 \right)   ]
\end{equation*}
and 
\begin{equation*}
\psi: [ x,y,z,t] \longrightarrow [y \left( t^2-y^2 \right) , -x(t^2 -y^2) + t \left( bt+ c(z-t)  \right) y,  \left(
  t-z \right) \left( t^2 - y^2 \right) ,t \left( t^2-y^2
  \right) ]
\end{equation*}
\begin{eqnarray*}
\kappa_\varphi =  \left( t+x \right) ^{4} \left( t-x \right) ^{4} = B_1^4 \,
 C_1^4, \qquad
\kappa_\psi =  \left( t+y \right) ^{4} \left( t-y \right) ^{4}
\end{eqnarray*}
The sequence of point we get from $p_0= [x,y,z,t]$ is
\begin{eqnarray}
&& p_1 = [ A_1, \, x\, B_1\,C_1,\, (z+t)\, B_1\,C_1, \, t\,B_1\,C_1]
  \nonumber \\ && p_2 = [ A_2\,B_1\,C_1,\, A_1\,B_2\,C_2,\,
    (z+2\,t)\,B_1\,C_1\,B_2\,C_2, \, t\,B_1\,C_1\,B_2\,C_2] \nonumber
  \\ && \dots \nonumber \\ && p_k = [ A_k\,B_{k-1}\,C_{k-1},\,
    A_{k-1}\,B_k\,C_k,\, (z+k\,t)\,B_{k-1}\,C_{k-1}\,B_k\,C_k,\,
    t\,B_{k-1}\,C_{k-1}\,B_k\,C_k] \label{stabledpII}
\end{eqnarray}
The recurrence on  $ A_k\,B_{k}\,C_{k}$ is given by the following constraints, 
generalising straightforwardly (\ref{hiromcmillan}):
\begin{eqnarray}
\boxed{
\begin{split}
%\begin{cases}
& A_k - t\, B_k\, C_{k} + {\bf B_{k+1} }\; C_{k-1} = 0, \quad A_k +
  t\, B_k\, C_k - B_{k-1} {\bf C_{k+1}} = 0 \\ & A_{k-1}\,{\bf
    B_{k+1}} \, {\bf C_{k+1}} + {\bf A_{k+1}} \, { B_{k-1}}\,
  C_{k-1}\,- t\,( b\,t +c \,(z+k\, t) )\, A_k\, B_k\, C_k = 0
%\end{cases}
\end{split}
}
\end{eqnarray}
The proof is similar to the one given in the previous section.
Setting here again $\Gamma_k = B_k\, C_k$ we get the `Hirota form'
found in~\cite{OhRaGrTa96}, extending over a string of length $3$.

We can use these relations to prove vanishing of the entropy, and
check that the various $A_k, B_k, C_k$ are the proper transforms of
$A_{k-1}, B_{k-1}, C_{k-1}$. The algebraic invariant has disappeared,
but the overall algebraic structure is essentially unchanged, compared
to the previous model, apart from one coefficient which became
non-constant.

\subsection{$qP_{VI}$}

The map $\varphi$, which is a discrete version of the Painlev\'e
equation $P_{VI}$, may be written as the composition three maps, taken
from equations (19,20,21) of~\cite{JiSa96}.
\begin{eqnarray*}
&\varphi (p_0) = \varphi( [x,y,z,t]) = \varphi_3 \cdot \varphi_2 \cdot
  \varphi_1( [x,y,z,t]) \qquad \mbox{with} \\ &\varphi_1(p_0) = [xy
    \left( x-ct \right) \left( x-dt \right) ,sh \left( x-az \right)
    \left( x-bz \right) {t}^{2},zy \left( x-ct \right) \left( x-dt
    \right) ,ty \left( x-ct \right) \left( x-dt \right) ]
  \\ &\varphi_2(p_0 ) = [cd \left( y-pz \right) \left( y-zr \right)
    {t}^{2},yx \left( -st+y \right) \left( y-ht \right) ,zx \left(
    -st+y \right) \left( y-ht \right) ,tx \left( -st+y \right) \left(
    y-ht \right) ] \\ &\varphi_3( p_0) = [x, y, q\, z , t], \qquad q =
  (cd \, pr)/( ab \, sh)
\end{eqnarray*}

The first iterates yield the following sequence of points, setting $\alpha=ab$, $\beta=cd$, $\sigma=sh$, and $\rho=pr$:
\begin{eqnarray*}
&&p_1 = [\alpha\beta\, y\, A_1\, B_1 \, E_1 \, F_1, \alpha\,\sigma^2
    \, xt^2\, C_1\, D_1\, G_1\, H_1, \beta\rho\, xy z\, C_1\, D_1\,
    E_1\, F_1, \alpha\sigma\, xyt\, C_1\, D_1\, E_1\, F_1]\\ &&p_2
  =[(\alpha\beta)^2\sigma \, xt^2\, A_2\, B_2 \, E_2 \, F_2,
    \alpha\beta\, \sigma^2 \, y\, C_2\, D_2\, G_2\, H_2,
    (\beta\rho)^2\, z\, C_2\, D_2\, E_2\, F_2, (\alpha\sigma)^2\, t\,
    C_2\, D_2\, E_2\, F_2]\ \\ &&p_3 = [(\alpha\beta)^3 \sigma^2\, y\,
    A_3\, B_3 \, E_3 \, F_3, \alpha\, \beta^2\sigma^2 \, x\, C_3\,
    D_3\, G_3\, H_3, (\beta\rho)^3\, xyz\, C_3\, D_3\, E_3\, F_3,
    (\alpha\sigma)^3\, xyt\, C_3\, D_3\, E_3\, F_3] \\ && p_4 =
  [(\alpha\beta)^4 \sigma^3\, x\, A_4\, B_4 \, E_4 \, F_4, \alpha\,
    \beta^3\sigma^2 \, yt^2\, C_4\, D_4\, G_4\, H_4, (\beta\rho)^4\,
    z\, C_4\, D_4\, E_4\, F_4, (\alpha\sigma)^4\, t\, C_4\, D_4\,
    E_4\, F_4] \\ && p_5 = [(\alpha\beta)^5 \sigma^4\, yt^2\, A_5\,
    B_5 \, E_5 \, F_5, \alpha\, \beta^4\sigma^2 \, x\, C_5\, D_5\,
    G_5\, H_5, (\beta\rho)^5\, xyz\, C_5\, D_5\, E_5\, F_5,
    (\alpha\sigma)^5\, xyt\, C_5\, D_5\, E_5\, F_5] \\ && p_6 =
  [(\alpha\beta)^6 \sigma^5\, x\, A_6\, B_6 \, E_6 \, F_6, \alpha\,
    \beta^5\sigma^2 \, y\, C_6\, D_6\, G_6\, H_6, (\beta\rho)^6\, z\,
    C_6\, D_6\, E_6\, F_6, (\alpha\sigma)^6\, t\, C_6\, D_6\, E_6\,
    F_6] \\ && \dots
\end{eqnarray*}

The form of $p_k$ is thus
\begin{eqnarray}
p_k= [(\alpha\beta)^k \sigma^{k-1} f_{1,k}\, A_k B_k E_k F_k, \alpha\,
  \beta^{k-1}\sigma^2 f_{2,k}\, C_k D_k G_k H_k,
  (\beta\rho)^kf_{3,k}\, C_k D_k E_k F_k, (\alpha\sigma)^kf_{4,k}\,
  C_k D_k E_k F_k]
\label{stableqpVI}
\end{eqnarray}
where the $f_{i,k}$ depend on the initial conditions, are such that
$f_{i,k}=f_{i,k+6}$, and can be read from the iterates given above.
They are summarised in the table
\begin{center}
\begin{tabular}{|c |c|c|c|c|}
\hline
    k  mod 6  &$ i=1$ & $i=2$ & $i=3$ & $i=4$ \\
\hline 
0  &$ x$ &$ y$ &$ z$ &$ t$ \\
%\hline 
1 &$ y$ & $xt^2$ & $xyz$ &$ xyt$ \\
%\hline 
2 & $xt^2$ &$ y$ &$ z $&$ t$ \\
%\hline 
3 &$ y$ &$ x$ &$ xyz$ &$ xyt$ \\
%\hline 
4  &$ x$ &$ yt^2$ &$ z$ &$ t$ \\
%\hline 
5  &$ yt^2$ &$ x$ &$xyz$ & $xyt$ \\
\hline 
\end{tabular}
\end{center}

The various factors $A,B,C,D,E,F,G,H$ verify simple recurrence
relations, which can be checked for the first few ones, and then
proved by recursion, as in the previous sections.

They read 
\begin{eqnarray}
\label{EFGHtoAB}
\boxed{
\begin{split}
%\begin{cases}
& p^{k-1}r^{k} \, \nu_{1,k} \, E_k F_k - sh\, \omega_{1,k} \, G_k H_k =
{\bf A_{k} }{B_{k-1}}
\\ 
& p^{k}r^{k-1}\, \nu_{1,k}\, E_k F_k - sh\,
\omega_{1,k}\, G_k H_k = A_{k-1} {\bf B_{k}}
\\ 
& (ab)^{k-1} s^{k-2} h^{k-1} \, \nu_{2,k} \, E_k F_k - (cd)^{k-1}\, 
\omega_{2,k} \, G_k H_k
= {\bf C_{k}} { D_{k}} 
\\ 
 & (ab)^{k-1} s^{k-1} h^{k-2} \,\nu_{2,k} \,
E_k F_k - (cd)^{k-1}\, \omega_{2,k}\, G_k H_k = C_{k-1} {\bf D_{k}}
%\end{cases}
\end{split}
}
\end{eqnarray}
 giving $A_k, B_k, C_k, D_k$ in terms of $A_{k-1}, B_{k-1}, E_k, F_k,
 G_k, H_k $, and
\begin{eqnarray}
\label{ABCDtoEF}
\boxed{
\begin{split}
%\begin{cases}
&  c^{k}d^{k-1} \, \lambda_{1,k} A_k B_k - sh \, \mu_{1,k} C_kD_k ={\bf
   E_{k+1}} F_k 
\\ 
& c^{k-1}d^{k} \, \lambda_{1,k} A_k B_k - sh \,
 \mu_{1,k} C_kD_k = E_{k} {\bf F_{k+1}} 
\\ 
& a^{k} b^{k-1} (sh)^{k-1} \,
 \lambda_{2,k} A_kB_k - (pr)^k \, \mu_{2,k} C_kD_k = {\bf G_{k+1} }{
   H_{k}} 
\\ 
& a^{k-1} b^{k} (sh)^{k-1} \, \lambda_{2,k} A_kB_k - (pr)^k
 \, \mu_{2,k} C_kD_k = G_{k} {\bf H_{k+1}}
%\end{cases}
\end{split}
}
\end{eqnarray}
giving $ E_{k+1}, F_{k+1}, G_{k+1}, H_{k+1} $ in terms of $A_k, B_k,
C_k, D_k$.  The coefficients $\lambda, \mu, \nu, \omega$ appearing in
the previous relations depend on the initial conditions and are given
by the following table:

\begin{center}
\begin{tabular}{|c|c|c|c|c|c|c|c|c|}
\hline
k mod 6 & $\nu_{1,k}$ & $\nu_{2,k}$ & $\omega_{1,k}$ & $\omega_{2,k}$& $\lambda_{1,k}$ & $\lambda_{2,k}$ & $\mu_{1,k}$ & $\mu_{2,k}$ \\
\hline
0 &$ z$ & $t$& $y$ & $y$ & $x$ & $x$ & $t$ & $z$
 \\
1 &$ yz$ & $y$ & $t^2$ & $t$ & $1$ & $1$ & $t$ & $xz$
\\
2 &$ z$ &$ t$  & $y$ & $y$ & $xt$ & $xt^2$ & $1$  & $z$
\\
3 &$ y$ &$ yt$ & $1$ & $1$ & $1$ & $1$ & $xt$ &$ xz$ 
\\
4 &$ z$ &$ 1$ &$ yt^2$ & $yt$ & $x$ & $x$ & $t$ &$ z$
\\
5 & $y$ & $yt$ &$ 1$ &$ 1$ & $t$ & $t^2$ & $x$ & $xz$ 
\\
 \hline
\end{tabular}
\end{center}

As in the previous case, equations (\ref{ABCDtoEF},\ref{EFGHtoAB}) can
be put in a  quadratic form by setting
\begin{eqnarray*}
U_k = A_k \, B_k,\; V_k = C_k \, D_k,\;  W_k = E_k\, F_k,\; T_k = G_k \, H_k,
\end{eqnarray*}
with coefficients depending on the initial conditions and on $k$. They
differ from the ones found in~\cite{OhRaGrTa96}.

Relations (\ref{EFGHtoAB}) and (\ref{ABCDtoEF}) define an iteration,
starting from $[A_1, B_1, C_1, D_1, E_1, F_1, G_1, H_1]$. The explicit
calculation of the first few iterates indicates that they are Laurent
polynomials in the initial conditions. If in addition the initial
conditions were produced by the action of the map $\varphi$ on some
$p_0= [x,y,z,t]$, then all resulting quantities are {\em polynomials}
in $x,y,z,t$. Relations (\ref{EFGHtoAB}) and (\ref{ABCDtoEF}), as well
as (\ref{stableqpVI}) may also be used to prove vanishing of the
entropy, because they provide an exact evaluation of the successive
degrees of the iterates of $\varphi$.

%%%%%%%%%%%%%%%%%%%%%%%%%%%%%%%%%%%%%%%%%%%%%%%%%%%%%%%%%%%%%%%%%%%%%%%%%%%%

\subsection{ A non integrable non confining map in $P_2$}
The model was proposed by F. Jaeger in relation to studies
of Bose-Meisner algebras (see for example~\cite{JaMaNo98}).  It is the
product of two birational involutions of $P_2$, constructed from a
projective linear map $L$ defined by the matrix
\begin{eqnarray*}
\left[ \begin {array}{ccc} 1&2&2\\ \noalign{\medskip}1&3&-4
\\ \noalign{\medskip}1&4&-5\end {array} \right] 
\end{eqnarray*}
and the fundamental involution
\begin{eqnarray*}
j:[x,y,z] \longrightarrow [ y\, z, z\,  z, x\, y].
\end{eqnarray*}
 Define 
\begin{eqnarray}
i = L^{-1} \cdot j \cdot L, \quad \varphi = i \cdot j, \quad \psi = j
\cdot i, \qquad \kappa_\varphi =\kappa_\psi = 25\, x\, y\, z.
\end{eqnarray}
It will appear that the pattern leads naturally to decompose $\varphi$
as the product
\begin{eqnarray*}
\varphi = ( L^{-1} \; j ) \cdot ( L \; j) 
\end{eqnarray*}
The sequence of iterates $p_0, \; q_1 = L^{-1}j (p_0), p_1 = Lj(q_1),
\dots$ we obtain  is
\begin{eqnarray*} 
  p_0 & = &  [x,y,z] \\
  q_1& = & [U_1, V_1, W_1] \\
  p_1& = & [A_1,\;  x\, B_1,\;  x \, C_1] \\
  q_2& = &[U_2,\; U_1 \,V_2,\; U_1\,W_2] \\
  p_2& = & [A_2,\; A_1 \,B_2,\; A_1\ , C_2] \\
  \dots \\
  q_k& = &[U_k, \; U_{k-1}\, V_k,\; U_{k-1}\, W_k] \\
  p_k& = & [A_k, \; A_{k-1}\, B_k,\; A_{k-1}\, C_k]
\end{eqnarray*}

The recurrence on the blocks $A,B,C,U,V,W$ take now the form
\begin{eqnarray}
\boxed{
\begin{split}
%\begin{cases}
  \label{hirojaeger_1}
&  A_{k-1}\, B_{k}\, C_{k}+ 2\, C_{k}\, A_{k}+ 2\, A_{k}\, B_{k} =
  5\,{\bf U_{k+1}} 
\\ 
& A_{k-1}\, B_{k}\, C_{k}+3\, C_{k}\, A_{k}-4\,
  A_{k}\, B_{k} = 5\, U_{k}\, {\bf V_{k+1}} 
\\ 
& A_{k-1}\, B_{k}\,
  C_{k}+4\, C_{k}\, A_{k}-5\, A_{k}\, B_{k} = 5\, U_{k}\, {\bf     W_{k+1}}
%\end{cases}
\end{split}
}
\end{eqnarray}

\begin{eqnarray}
\boxed{
\begin{split}
%\begin{cases}
\label{hirojaeger_2}
& U_k\, V_{k+1}\, W_{k+1}+18\; W_{k+1}\, U_{k+1}-14\, U_{k+1}\,V_{k+1}
={\bf A_{k+1}}, 
\\ 
& U_{k}\, V_{k+1}\, W_{k+1}-7\; W_{k+1}\, U_{k+1}+6\,
U_{k+1}\, V_{k+1} = A_{k}\,{\bf B_{k+1}}, 
\\
& U_{k}\, V_{k+1}\,
W_{k+1}-2\; W_{k+1}\, U_{k+1}+U_{k+1}\, V_{k+1} = A_{k}\,{\bf C_{k+1}}
%\end{cases}
\end{split}
}
\end{eqnarray}

The recurrence relations we get here are not quadratic.  The main
point is that in (\ref{hirojaeger_1}) and (\ref{hirojaeger_2}),
$U_{k+1}$ and $A_{k+1}$ have polynomial expressions. This reflects the
singularity structure of the map: in the iteration process, the lines
$\{ y=0\}$ and $\{ z=0 \}$ are blown down to points which whose images
never meets any singularity, while the line $\{ x=0\}$ goes to
$[1,1,1]$, and then $[1,0,0]$ which is singular. Relations
(\ref{hirojaeger_1}) and (\ref{hirojaeger_2}) allow us to calculate
exactly the entropy $\epsilon = \log ( (3+\sqrt{5})/2 )$ of $\varphi$.

%%%%%%%%%%%%%%%%%%%%%%%%%%%%%%%%%%%%%%%%%%%%%%%%%%%%%%%%%%%%%%%%%%%%%%%%%%%%

\subsection{ A confining non integrable map in $P_2$}
We briefly mention here what has become a prototype of confining but
chaotic map in two dimension, described in~\cite{HiVi98}.
\begin{eqnarray*}
\varphi: [ x,y,z] \longrightarrow [{x}^{3}+a{z}^{3}-y{x}^{2},{x}^{3},{x}^{2}z]
\end{eqnarray*}
coming from the simple order 2 recurrence
\begin{eqnarray*}
u_{n+1} + u_{n-1} = u_n +\frac{a}{u_n^2}
\end{eqnarray*}
Here
\begin{eqnarray*}
\kappa_\varphi = x^3, \qquad \kappa_\psi = y^3
\end{eqnarray*}

This map has been shown to have positive entropy by various methods,
among which the construction of a rational surface over $P_2$ where
the singularities are resolved~\cite{Ta01c}. The lift of the map to
the Picard group of this variety is a linear map whose maximal
eigenvalues gives the entropy.

The generic iterate has the form 
\begin{eqnarray}
p_k=[  A_{k-3}^3\, A_k,  A_{k-4}\, A_{k-1}^3, z\, A_{k-3}^2\, A_{k-2}^2\, A_{k-1}^2 ]
\label{stablehv}
\end{eqnarray}
and the recurrence relation between the blocks $A$ becomes
\begin{eqnarray}
\label{hirohv}
\boxed{
A_k^3\, A_{k-3}^3+a\, z^3\, A_{k-1}^6\, A_{k-2}^6 - A_{k-1}^3\,
A_{k-4}\, A_{k}^2 =\, A_{k-3}^2\, A_{k-2}^3\, {\bf A_{k+1}}
}
\end{eqnarray}
which is equ (4.6) of~\cite{Ho07}, extending over a string of length
$6$.  This relation is not quadratic nor multilinear, but it allows to
prove (\ref{stablehv}) with the same type of argument as in the
previous sections, providing the recurrence condition on the degrees
of the iterates of $\varphi$, and the value of the entropy $\epsilon=
(3+\sqrt{5})/2$.

%%%%%%%%%%%%%%%%%%%%%%%%%%%%%%%%%%%%%%%%%%%%%%%%%%%%%%%%%%%%%%%%%%%%%%%%%%%%

\subsection{ Another confining non integrable map in $P_2$}

Another interesting confining non integrable map described
in~\cite{Vi08b}, eq 29,  was also examined for $a=0$ in~\cite{Ho07}.
The map comes from the order $2$ recurrence
\begin{eqnarray*}
u_{n+1} \cdot u_{n-1} = u_n+ \frac{1}{u_n} +a
\end{eqnarray*}
\begin{eqnarray*}
&& \varphi: [x,y,z] \rightarrow [z \left( {x}^{2}+{z}^{2}+axz \right)
    ,y{x}^{2},xyz], \\
&& \psi: [x,y,z] \rightarrow [{y}^{2}x,z \left(
    {y}^{2}+{z}^{2}+azy \right) ,xyz].  \\ && 
\kappa_\varphi =  {x}^{3}{y}^{2}z \, (x+ z\, (a+\alpha)/2 ) \;  (x+ z\, (a-\alpha)/2 ) =   {x}^{3}{y}^{2}z \, B_1 \, C_1
 , \\
&&
  \kappa_\psi = {x}^{2}{y}^{3}z \, (y+ z\, (a+\alpha)/2 )  \; (y+ z\, (a-\alpha)/2 )  \end{eqnarray*}
with  $\alpha = \sqrt{a^2 -4}$.

By blowing up $18$ points, we may define a rational surface over $P_2$
where the lift of the map becomes a diffeomorphism, and the entropy is
given as the logarithm of the inverse of the root of $
{s}^{4}-{s}^{3}-2\,{s}^{2}-s+1 =0$ of smallest modulus ($\simeq
\log 2.08102$).

The iterates of $\varphi$ are\footnote{From $A_1=B_1 \, C_1$, the
  blocks $A_k$ are products of two blocks, which we do not write for
  simplicity}
\begin{eqnarray*}
&&p_0 = [x,y,z]\\
&&p_1 = [z\, A_1,x^2\, y,x\, y\, z]\\
&&p_2 = [z\, A_2, x\, A_1^2 ,x^2 \, y A_1]\\
&&p_3 = [x\, y\, A_3, z^2 \, A_1\, A_2^2  ,\, x^2 y\, z\, A_1^2\, A_2]\\
&&p_4 = [x^2 y \, A_1 A_4, z \, A_2\, A_3^2 , x\ z^2 \, A_1^2 \, A_2^2\,  A_3]\\
&&p_5 = [z \, A_1^2 \, A_2\, A_5,x\, y^2\, A_3\, A_4^2,y\, z^2 A_1\, A_2^2\, A_3^2\, A_4]\\
&&p_6 =  [z^2 \, A_2^2 \, A_3\, A_6,x\, y\,A_1\, A_4\, A_5^2,\, x\,y^2\, z A_2\, A_3^2\, A_4^2\, A_5]\\
&& \dots
\end{eqnarray*}
The form stabilises into
\begin{eqnarray}
p_k = [g_{1,k} \, A_{k-4}^2 \, A_{k-3}\, A_k, g_{2,k} \,A_{k-5}\, A_{k-2}\, A_{k-1}^2,\, g_{3,k} \, A_{k-4}\, A_{k-3}^2\, A_{k-2}^2\, A_{k-1}].
\label{stablecmv}
\end{eqnarray}
The coefficients $g_{i,k}$ depend on the initial condition
$\{x,y,z\}$, and verify $g_{i,k+9} = g_{i,k}$. They are given by:
\begin{center}
 \begin{tabular}{|c |c|c|c|}
\hline
    k  mod 9  &$ i=1$ & $i=2$ & $i=3$  \\
\hline 
0  &  $   x   $  &  $   y   $ &  $  z    $  \\
%\hline 
1  &  $   z   $  &  $ x^2y     $ &  $   xyz   $  \\
%\hline 
2  &  $ z     $  &  $  x    $ &  $ x^2y     $  \\
%\hline 
3  &  $   xy   $  &  $  z^2    $ &  $ x^2yz     $  \\
%\hline 
4  &  $   x^2y   $  &  $  z    $ &  $    xz^2  $  \\
%\hline 
5  &  $    z  $  &  $  xy^2    $ &  $ yz^2     $  \\
%\hline 
6  &  $  z^2    $  &  $ xy     $ &  $ xy^2z     $  \\
%\hline 
7  &  $  y    $  &  $   z   $ &  $ xy^2     $  \\
%\hline 
8  &  $   xy^2   $  &  $ z     $ &  $ xyz     $  \\
\hline 
\end{tabular}
\end{center}
Notice that the presence of the factors $g_{i,k}$ shows that the
 coordinate planes appear periodically in the iterates.

The recurrence relations between the $A$'s also have coefficients
depending on the initial conditions, and on $k$ in a periodic way
(period 9). 
\begin{eqnarray}
\label{hirocmv}
\boxed{
\lambda_k^2 \, A_{k-4}^2 \,A_{k}^2 +\mu_k^2 \, A_{k-3}^2\, A_{k-2}^4\,
A_{k-1}^2 + a\, \lambda_k \mu_k \, A_{k-4}\, A_{k-3}\, A_{k-2}^2 \,
A_{k-1}\, A_{k}\, = A_{k-5}\, {\bf A_{k+1}}
}
 \end{eqnarray}
The coefficients are given by the following table
\begin{center}
 \begin{tabular}{|c |c|c|}
\hline
    k  mod 9  &$ \lambda$ & $\mu$   \\
\hline 
0  &  $   x   $  &  $   z   $  \\
%\hline 
1  &  $   1   $  &  $ xy     $  \\
%\hline 
2  &  $ z    $  &  $  x^2y    $   \\
%\hline 
3  &  $  1   $  &  $  xz    $  \\
%\hline 
4  &  $   xy   $  &  $  z^2    $    \\
%\hline 
5  &  $    1  $  &  $  yz    $    \\
%\hline 
6  &  $  z   $  &  $ xy^2     $   \\
%\hline 
7  &  $  1    $  &  $   xy  $   \\
%\hline 
8  &  $   y   $  &  $ z     $   \\
\hline 
\end{tabular}
\end{center}

The recurrence relation (\ref{hirocmv}) differs from the one given for
$a=0$ in~\cite{Ho07}, since it depends on the order $k$. The explicit
calculation of the first iterations indicates that it verifies the
Laurent property for arbitrary $x,y,z$ and $a$.

%%%%%%%%%%%%%%%%%%%%%%%%%%%%%%%%%%%%%%%%%%%%%%%%%%%%%%%%%%%%%%%%%%%%%%%%%%%%

\subsection{An integrable map in $P_3$: N=3 Periodic 
Volterra}

\begin{eqnarray*}
\varphi:= [x,y,z,t] \longrightarrow [x',y',z',t'] \qquad \mbox{ with }
\end{eqnarray*}
\begin{eqnarray}
x' & = & - x \left( t^2 + 2\, e \, t \, (y-z) -e^2 x^2 +e^2 ( y+z)^2
\right) \label{surat_x} \\ t'& = & -t \left( t^2 + e^2 ( x^2 + y^2 +
z^2 - 2xy-2yz - 2xz) \right)
\end{eqnarray}

$y'$ and $z'$ being obtained from (\ref{surat_x}) by circular
permutations of $x,y,z$.

The map comes from a discretisation of a continuous integrable system,
and is known to have two algebraic invariants~\cite{PePfSu11}. 

Starting from $p_0= [A_0, B_0, C_0, D_0]$ we get a sequence of points
of the form:
\begin{eqnarray*}
p1 & =& [A_0\,A_1, B_0\,B_1, C_0\,C_1, D_0\,D_1]\\
p2 & =& [A_0\,A_1\,A_2, B_0\,B_1\,B_2, C_0\,C_1\,C_2, D_0\,A_1\,B_1\,C_1]\\
p3 & =& [A_0\,A_1\,A_2\,A_3, B_0\,B_1\,B_2\,B_3, C_0\,C_1\,C_2\,C_3, D_0\,D_1\,B_2\,A_2\,C_2]\\
p4 & =& [A_0\,A_2\,A_3\,A_4, B_0\,B_2\,B_3\,B_4, C_0\,C_2\,C_3\,C_4, D_0\,B_3\,A_3\,C_3]\\
p5 & =& [A_0\,A_3\,A_4\,A_5, B_0\,B_3\,B_4\,B_5, C_0\,C_3\,C_4\,C_5, D_0\,D_1\,B_4\,A_4\,C_4] \\
& \dots &
\end{eqnarray*}
The form of the iterates stabilises after three steps, with as slight
difference in the structure of the last component between odd and even
order.

Denoting the $k$th iterate $p_k$ as $[X_k, Y_k,Z_k,T_k]$ with
\begin{eqnarray}
\begin{cases}
\label{stablevolte}
& X_{k} = A_0\,A_{k-2}A_{k-1}A_k, \qquad Y_{k} = B_0\,B_{k-2}B_{k-1}B_{k}, \qquad
 Z_{k} = C_0\,C_{k-2}C_{k-1}C_{k} \\
& T_{2m+1} = D_0 D_1 A_{2m} B_{2m} C_{2m},  \qquad
T_{2m+2} = D_0 A_{2m+1} B_{2m+1} C_{2m+1},
\end{cases}
\end{eqnarray}
the various $A,B,C,D$'s verify sets of constraints of the form
\begin{eqnarray}
\label{ideal_surat}
\begin{cases}
\label{ideal_surat_all_1}
 &  D_0 ( X_{k} + Y_{k} + Z_{k})  - ( A_0+B_0+C_0)\, T_{k}  = 0   \\
\label{ideal_surat_all_2}
 & T_{k} + e ( \pm X_{k} \pm Y_{k} \pm Z_{k}) - \left( D_0 + e( \pm
A_0 \pm B_0\pm C_0) \right)\, \alpha_{k-1} \, \beta_{k}\, \gamma_{k+1}
= 0
\end{cases}
\end{eqnarray}
where $\{ \alpha, \beta, \gamma\}$ is some permutation of $\{A, B, C\}$.

These constraints, extending over strings of successive points of
length $3$, are responsible for the factorisation properties: they
define various ideals, and the factorisations take place in the the
algebra generated by the $A,B,C,\dots$ {\em quotiented by these
  ideals}. Moreover the constraints are conserved by the
evolution. The various $\pm$ signs in (\ref{ideal_surat}) depend on
$k$ in a periodic way (period 3).

A typical example of these relations is (for $k=3$):
\begin{eqnarray*}
\begin{cases}
& D_0\, ( X_3 + Y_3 + Z_3)  - ( A_0+B_0+C_0)\, T_3  = 0 \\
& T_3 + e ( -X_3+Y_3+Z_3) -  ( D_0 -e A_0+eB_0+eC_0)\, A_2\, B_3\, C_1 = 0 \\
& T_3 + e ( X_3-Y_3+Z_3) -  ( D_0 +  eA_0-eB_0+eC_0)\, A_1\, B_2\, C_3 = 0  \\
& T_3 + e ( X_3+Y_3-Z_3) -  ( D_0 + eA_0+eB_0-eC_0) \,A_3 \,B_1\, C_2 = 0  \\
& T_3 + e ( -X_3-Y_3+Z_3) -  ( D_0 -eA_0-eB_0+eC_0) \,A_1\, B_3\, C_2 = 0  \\
& T_3 + e ( -X_3+Y_3-Z_3) -  ( D_0 -e A_0+eB_0-eC_0)\, A_3\, B_2\, C_1 = 0  \\
& T_3 + e ( X_3-Y_3-Z_3) -  ( D_0 + eA_0-eB_0-eC_0) \,A_2\, B_1\, C_3 = 0  \\
\end{cases}
\end{eqnarray*}
  The set of constraints is invariant by circular permutation $ A
  \rightarrow B\rightarrow C \rightarrow A $. It cannot be written
  solely in term of the components $X,Y,Z,T$. It cannot be solved
  straightforwardly for any set $A_k, B_k,C_k$'s, because it is then
  over-determined.

On the other hand, writing that $[X_{k+1},Y_{k+1},Z_{k+1},T_{k+1}]$ is
the image of $[X_k,Y_k,Z_k,T_k]$ by $\varphi$ yields algebraic
equations for $\{ A_{k+1}, B_{k+1}, C_{k+1}\}$, which one can solve
rationally.

The effect of the set of constraints
(\ref{ideal_surat}) is that these
expressions can be simplified. The factors of $\{ A_{k+1}, B_{k+1},
C_{k+1}\}$, reduce to monomials in  $ A_{k-1}, B_{k-1},C_{k-1},
A_{k-2}, B_{k-2},C_{k-2} $ and possibly $D_1$ for odd $k$.

The simplified relations defining the iteration read:
\begin{eqnarray}
\label{hirota4}
\boxed{
%\begin{cases}
\begin{split}
& T_{k}^2+ 2eT_{k} (Y_{k}-Z_{k}) + e^2(Y_{k}+Z_{k})^2 - e^2 X_{k}^2 +  \Delta_k {\bf A_{k+1}}B_{k-2}C_{k-2}A_{k-1}B_{k-1}C_{k-1} = 0 \\
& T_{k}^2+ 2eT_{k} (Z_{k}-X_{k}) +e^2(Z_{k}+ X_{k})^2 - e^2 Y_{k}^2 + \Delta_k A_{k-2}{\bf B_{k+1}}C_{k-2}A_{k-1}B_{k-1}C_{k-1} = 0 \\
& T_{k}^2+ 2eT_{k} (X_{k}-Y_{k}) + e^2(X_{k}+Y_{k})^2 - e^2 Z_{k}^2+ \Delta_k A_{k-2}B_{k-2}{\bf C_{k+1}}A_{k-1}B_{k-1}C_{k-1} = 0 
%\end{cases}
\end{split}
}
\end{eqnarray}
with $\Delta_k = D_1$ for odd $k$ and $\Delta_k=1$ for even $k$.

These equations are linear in $\{A_{k+1}, B_{k+1}, C_{k+1}\}$.  They
tell us that the factor
\begin{eqnarray*}
f_k =  \Delta_k \, A_{k-2}\, B_{k-2} \, C_{k-2}\, A_{k-1}\, B_{k-1}\, C_{k-1}
\end{eqnarray*}
goes away from the homogeneous coordinates when calculating $p_{k+1}$
as $\varphi(p_k)$. They extend over a {\em string of successive points
  of length $4$.}  They are the generalisation of the Hirota bilinear
formalism for the map under consideration, but they are not quadratic
anymore.  They do not have the Laurent property.

{\em Thanks to the relations
  (\ref{ideal_surat}), their solution in
  $\{A_{k+1}, B_{k+1}, C_{k+1}\}$ is polynomial in terms of the
  initial conditions $\{A_{0}, B_{0}, C_{0}, D_{0}\}$}, and it is
possible to show that {\em $A_{k+1}, B_{k+1}, C_{k+1}$ are
  the proper transforms of $A_{k}, B_{k}, C_{k}$}. In other
words, the $A, B, C$'s do not  factorise.

%%%%%%%%%%%%%%%%%%%%%%%%%%%%%%%%%%%%%%

\subsection{A linearisable map}

Linearisable recurrence are known to have special singularity
structure. As an example we can take the one studied
in~\cite{AbHaHe00}, where it was shown to be non-confining, but
integrable.

The recurrence is
\begin{eqnarray*}
u_{n+1}=u_n+\frac{u_n-u_{n-1}}{1+u_n-u_{n-1}}
\end{eqnarray*}
so that 
\begin{eqnarray*}
&& \varphi: [x,y,z] \longrightarrow [2\,xz+{x}^{2}-xy-yz,x \left( x-y+z \right) , \left( x-y +z\right) z] \\
&& \psi:  [x,y,z] \longrightarrow [y \left( x-y-z \right) ,xz+xy-2\,yz-{y}^{2},z \left( x-y-z \right) ]
\\
&&\kappa_\varphi = z^2 \; (x-y+z)=z^2 \; B1, \qquad \kappa_\psi =  z^2 \; (x-y-z)
\end{eqnarray*}

The iterates $p_k$ take the form
\begin{eqnarray*}
p_k = [ A_k, A_{k-1} \; B_k, z\, B_1 \, B_2 \, B_3 \dots \, B_k]
\end{eqnarray*}
with 
\begin{eqnarray*}
\varphi^* ( B_k ) = z \; B_{k+1}
\end{eqnarray*}

There is regularity of the pattern, but the number of factors
increases with $k$. This is related to the fact that at each step, one
more factor $B_1$ appears, and is at the origin of the low (linear)
growth of the degrees of the iterates.

The recurrence relations between blocks read
\begin{eqnarray}
\label{hiroabhahe}
\boxed{
\begin{split}
%\begin{cases}
& z\,  B_1\, B_2 \dots \, B_{k-1} \,  ( 2 \, A_k\, B_k -  A_{k-1}\,  B_k^2 - {\bf A_{k+1}}) + A_k ( A_k  -  A_{k-1} B_k) = 0\\
& z\,  B_1\, B_2 \dots \, B_{k-1} \, (B_k - {\bf B_{k+1}}) +  A_k  -  A_{k-1}\, B_k = 0
%\end{cases}
\end{split}
}
\end{eqnarray}

The linear growth of the degrees can be read from the previous relations.

%%%%%%%%%%%%%%%%%%%%%%%%%%%%%%%%%%%%%%%%%%%%%%%%%%%%%%%%%%%%%%%%%%%%%%%%%%%%

\subsection{An unruly  model}
\label{unruly}

We know of maps for which the sequence of degrees does not verify any
finite recurrence relation. Although this does not prevent their
entropy from being the logarithm of an algebraic integer, it will
prevent the existence of the Hirota like forms we have seen in the
previous cases.

A simple example was found in~\cite{HaPr05}.  It is a monomial map
in three dimensions:
\begin{eqnarray*}
&& \varphi: [x,y,z,t] \longrightarrow [ty,tz,{x}^{2},tx] ,\qquad
\psi: [x,y,z,t] \longrightarrow  [tz,xz,yz,{t}^{2}] \\
&& \kappa_\varphi =  x^2 t,\qquad
\kappa_\psi = z t^2 
\end{eqnarray*}
This map  has the peculiarity that the entropies of $\varphi$ and
of $\psi$ differ~\cite{HaPr05}.

The structure of the iterates is simple, since they are all written in
term of the coordinate planes
\begin{eqnarray*}
p_k = [ x^{\delta_1^x} \,  y^{\delta_1^y} \, z^{\delta_1^z} \, t^{\delta_1^t},
 x^{\delta_2^x} \,  y^{\delta_2^y} \, z^{\delta_2^z} \, t^{\delta_2^t},
 x^{\delta_3^x} \,  y^{\delta_3^y} \, z^{\delta_3^z} \, t^{\delta_3^t},
 x^{\delta_4^x} \,  y^{\delta_4^y} \, z^{\delta_4^z} \, t^{\delta_4^t}]
\end{eqnarray*}
for some powers $\delta_*^*$. For $\psi$, the form of the iterates
stabilises and sequence of degrees verifies a finite recurrence
relation, but this is not the case for $\varphi$. The peculiarity of
the model is that the singularity structure is such that the sequences
of proper transforms which are at the basis of the observation we made
for all other examples do not appear here.  This model is a limiting
case to keep in mind for further developments.

%%%%%%%%%%%%%%%%%%%%%%%%%%%%%%%%%%%%%%%%%%%%%%%%%%%%%%%%%%%%%%%%%%%%%%%%%%%%

\subsection{A delay-differential / differential difference 
equation}

Consider the following equation:
\begin{eqnarray}
\label{qcs}
a \; u(t) - b \; \partial_t{u}(t) = u(t) \; \left(  u(t+1) - u(t-1) \right) 
\end{eqnarray}
where $\partial_t $ means time derivative.

Equation (\ref{qcs}) was obtained in~\cite{QuCaSa92} by a non trivial
reduction of a semi-discrete equation~\cite{Ma74}.  This equation is a
delay difference equation of which the entropy has been evaluated
in~\cite{Vi14}, and found to be vanishing.  One may equivalently
consider the differential difference equation, or recurrence of order
two defined on functional space:
\begin{eqnarray}
\label{qcs_alter}
a \; u_n(t) - b \; \partial_t{u}_n(t) = u_n(t) \; \left( u_{n+1}(t) -
u_{n-1}(t) \right)
\end{eqnarray}

 The maps $\varphi$ and  $\psi$  associated to these equations are:
\begin{eqnarray}
\label{phi}
\varphi: [x,y,z] &\longrightarrow & [ a \, x z - b\, ( x' z - x z') +
  x y, \; x^2,\; x z] \\
\label{psi}
\psi : [x,y,z] & \longrightarrow &[ y^2, \; - a\, y z + b\, ( y' z - y
  z') + x y , \; y z ]
\end{eqnarray}
were prime ($'$) means derivative. Here $x,y,z$ should be considered
as a container for the infinite sequences $[ x(t), x'(t), x''(t),
  \dots]$, $ [ y(t), y'(t), y''(t), \dots]$, and $[ z(t), z'(t),
  z''(t), \dots]$.

For this map 
\begin{eqnarray*}
\kappa_\varphi ([x,y,z])= x^3 , \qquad \kappa_\psi([x,y,z])= y^3.
\end{eqnarray*}
and we get the following form for the first iterates starting from $p_0$
\begin{eqnarray}
p_0 &=& [ A_0, B_0 , C_0] 
\nonumber \\
p_1 & = & [\; A_1, \; A_0^2, \; A_0 C_0 \; ]\nonumber \\
p_2 & = & [ \; A_2  , \; A_1^2   , \; A_0 A_1 C_0 \;  ]\nonumber \\
p_3 & = & [\;  A_0^2 A_3 ,\;  A_2^2  , \; A_0  A_1 A_2 C_0\;  ]\nonumber \\
p_4 & = & [\;  A_1^2 A_4,  \;  A_0 A_3^2, \; A_1 A_2 A_3 C_0 \;  ]\nonumber \\
\dots\nonumber \\
p_k & = & [\;  A_{k-3}^2 \; A_k,  \;  A_{k-4} \; A_{k-1}^2, \; A_{k-3} \; A_{k-2} A_{k-1} \; C_0 \; ] \label{stabledelaydiff} 
 \end{eqnarray}

The recurrence for $A_k$ reads
\begin{eqnarray}
\label{hirodelay}
\boxed{
a\, { C_0}\,{ A_{k}}\,{ A_{k+1}}\,{ A_{k+2}}\,{ A_{k+3}}+{ A_0}\,{
  A_{k+3}}\, {{ A_{k+2}^2}}+c\, {{ A_{k}^2}}{{ A_{k+3}^2}}\,
{\partial_t} \left( {\frac {{ C_0}\,{ A_{k+1}}\,{ A_{k+2}}}{{
      A_{k}}\,{ A_{k+3}}}} \right) = { A_{k}}\,{{ A_{k+1}^2}}\,{{\bf
    A_{k+4}}}
}
\end{eqnarray}
This relation extends over a string of length $5$.  Again, although
given as a fraction, $A_{k+4}$ is a differential polynomial in the
initial conditions.

The proof of relation (\ref{hirodelay}) is done by recursion. It is
verified for $ k=0$ and $k=1$. We moreover know that the pullback of
any $A_k$ by $\varphi$ is of the form $A_0^n A_{k+1}$,
it is easy to show the validity of (\ref{hirodelay}) for $k+1$.  In
particular, one finds that $ A_{k+1}$ is the proper transform of $
A_{k}$ and moreover
\begin{eqnarray*}
 \varphi^* \left( {\frac {{C_0}\,{ A_{k+1}}\,{ A_{k+2}}}{{ A_{k}}\,{
       A_{k+3}}}} \right) = {\frac {{ C_0}\,{ A_{k+2}}\,{ A_{k+3}}}{{
       A_{k+1}}\,{ A_{k+4}}}},
\end{eqnarray*}

so that the pullback of the derivative term appearing in
(\ref{hirodelay}) does not contain factors $A_0$.  We also get
relations on the various degrees $\delta(A_k)$:
\begin{eqnarray*}
1+\delta(A_k)  +\delta(A_{k+1})+\delta(A_{k+2})+\delta(A_{k+3})& =&
1+2\, \delta(A_{k+2})  +\delta(A_{k+3}) \\
& = & \delta(A_k)  + 2\, \delta(A_{k+1})+\delta(A_{k+4}).
\end{eqnarray*}
One then easily proves the result on the sequence of degrees of $p_n$
given in~\cite{Vi14}
\begin{eqnarray*}
 \delta(p_n)  = \frac {1}{8} \; \bigg(  6 \; n^2 +9 - (-1)^n \bigg)
\end{eqnarray*}
ensuring the vanishing of the entropy. 

 One could always question the notion of integrability for
 differential-difference equations, and even more for delay-difference
 equations, but the vanishing of the algebraic entropy is a very
 strong structural constraint on the equation.

%%%%%%%%%%%%%%%%%%%%%%%%%%%%%%%%%%%%%%%%%%%%%%%%%%%%%%%%%%%%%%%%%%%%%%%%%%%%

\subsection{An integrable lattice map: $Q_V$}
\label{q5}

The model, introduced in~\cite{Vi06,Vi08} is defined on a plane square
lattice, by a multilinear relations between the values of an unknown
function $u_{n,m}$, $n\in Z, m\in Z$. It interpolates between the
various models of the Adler-Bobenko-Suris
list~\cite{AdBoSu03,AdBoSu04,AdBoSu07}, and has seven free
parameters. The integrability of the model was originally based on the
evaluation of its algebraic entropy, which vanishes. Subsequently this
model was shown to have an infinite set of symmetries, implemented by
two recursion operators related by a elliptic
condition~\cite{MiWaXe11,MiWa11}.

The elementary cell of the lattice, written with the usual convention
$u_{n,m}=u,\; u_{n+1,m}=u_1, \; u_{n,m+1}=u_2, \; u_{n+1,m+1}=u_{12}$
looks like
\vskip -1truecm
\begin{center}
\setlength{\unitlength}{1300sp}%
\begingroup\makeatletter\ifx\SetFigFont\undefined%
\gdef\SetFigFont#1#2#3#4#5{%
  \reset@font\fontsize{#1}{#2pt}%
  \fontfamily{#3}\fontseries{#4}\fontshape{#5}%
  \selectfont}%
\fi\endgroup%
\begin{picture}(6024,6024)(2989,-6973)
{\color[rgb]{0,0,0}\thinlines
\put(4201,-2161){\circle*{212}}
}%
{\color[rgb]{0,0,0}\put(7801,-2161){\circle*{212}}
}%
{\color[rgb]{0,0,0}\put(4201,-5761){\circle*{212}}
}%
{\color[rgb]{0,0,0}\put(7801,-5761){\circle*{212}}
}%
{\color[rgb]{0,0,0}\put(4201,-961){\line( 0,-1){6000}}
}%
{\color[rgb]{0,0,0}\put(7801,-961){\line( 0,-1){6000}}
}%
{\color[rgb]{0,0,0}\put(3001,-2161){\line( 1, 0){6000}}
}%
{\color[rgb]{0,0,0}\put(3001,-5761){\line( 1, 0){6000}}
}%
\put(8101,-6400){\makebox(0,0)[lb]{\smash{{\SetFigFont{20}{24.0}{\rmdefault}{\mddefault}{\updefault}{\color[rgb]{0,0,0}$u_{1}$}%
}}}}
\put(3000,-1936){\makebox(0,0)[lb]{\smash{{\SetFigFont{20}{24.0}{\rmdefault}{\mddefault}{\updefault}{\color[rgb]{0,0,0}$u_{2}$}%
}}}}
\put(3200,-6400){\makebox(0,0)[lb]{\smash{{\SetFigFont{20}{24.0}{\rmdefault}{\mddefault}{\updefault}{\color[rgb]{0,0,0}$u$}%
}}}}
\put(7951,-1936){\makebox(0,0)[lb]{\smash{{\SetFigFont{20}{24.0}{\rmdefault}{\mddefault}{\updefault}{\color[rgb]{0,0,0}$u_{12}$}%
}}}}
\end{picture}%
\end{center}
and the local relation defining the model reads:
\begin{eqnarray}
&& a_{{1}}\,\,u\, {\it u_1}{\it u_2}{\it u_{12}}+ a_{{2}} \, \left(u\,
  {\it u_1}\, {\it u_2}+u\, { \it u_2}\,{\it u_{12}}+u \,{\it u_1}\,
  {\it u_{12}}+{\it u_1}\,{\it u_2}\,{\it u_{12}} \right) + a_{{3}} \,
  \left( u\, {\it u_1}+{\it u_2}\,{\it u_{12}} \right) \nonumber \\ &&
  + a_{{4}}\, \left({\it u_1}\, {\it u_2}\, + u\, {\it u_{12}}
  \right)+ a_{{5}} \, \left(u\, {\it u_2 } +{\it u_1} \, {\it u_{12}}
  \right) + a_{{6}}\, \left(u+ {\it u_1}+{\it u_2}+{ \it u_{12}}
  \right) +a_{{7}} = 0
\end{eqnarray}
The previous relation being multilinear, it is possible to calculate
any of the corner variables in term of the other three.  On each cell
set $\varphi_{n,m}: u \longrightarrow u_{12}$ and $\psi_{n,m}: u_{12}
\longrightarrow u $.  

In order to define an evolution we need to specify initial
conditions. We choose to give initial conditions on two adjacent
diagonals, labelled $-1$ and $0$, and use the local condition to fill
the entire lattice. Points on the diagonal $k$ have coordinates $\{
n,m\}$ with $n+m=k$. We may then define a map $\varphi$ from diagonal
$k$ to $k+2$, and $\psi$ from diagonal $k+2$ to $k$ (straight arrows
in Figure~1), for which the values on diagonal $k+1$ enter as
parameters.  Although the space of initial conditions is infinite, a
point $\{n,m\}$, 'sees' only a finite number of initial points on the
diagonals $-1$ and $0$. We projectivise the system by turning the
space of values at each point in to a projective line $P_1$, with
homogeneous coordinates $[X_{n,m}, Y_{n,m}]$ so that $u_{n,m} =
X_{n,m}/Y_{n,m}$, and writing only polynomial expressions, keeping in
mind that any common factor to $X$ and $Y$ ought to be removed.

\begin{center}
\includegraphics{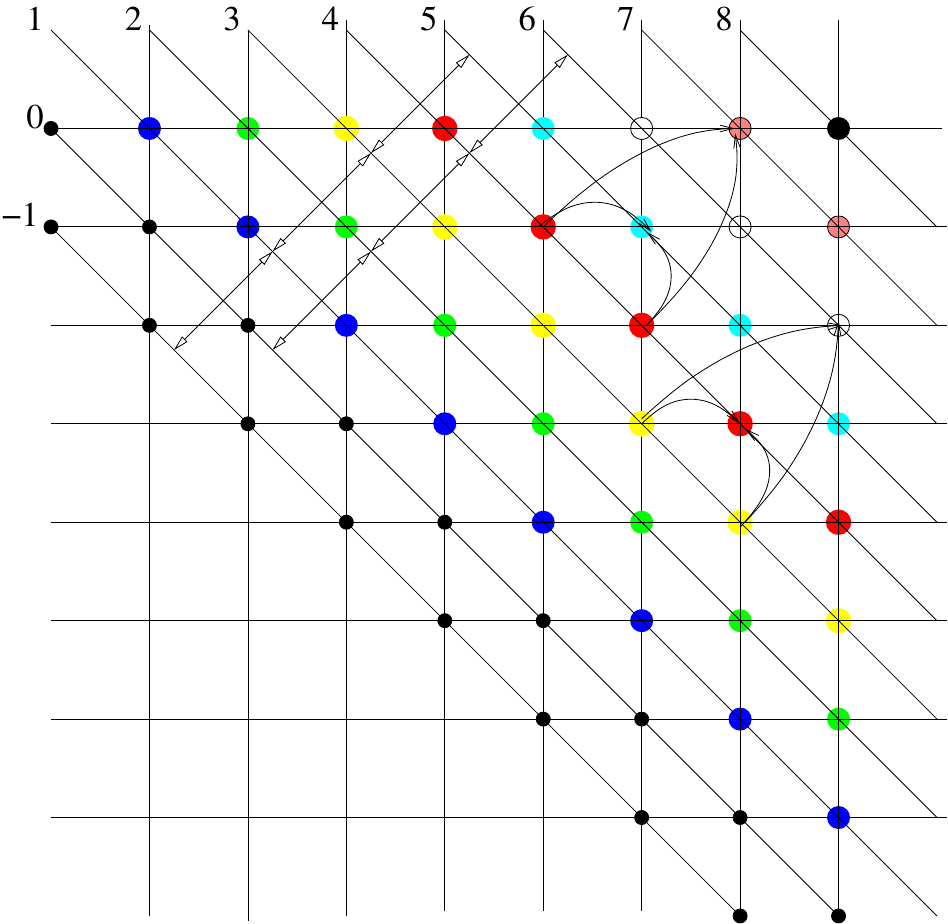}\par
Figure 1: Initial conditions and (north-east) evolution
\end{center}

We know from~\cite{Vi08} that the drop of the degrees of the
successive iterates is intimately related to one of the biquadratics
given in~\cite{AdBoSu03}. This biquadratic is nothing but the
multiplier $\kappa_{{n,m}} \simeq \psi_{n,m}\cdot\varphi_{n,m}$
calculated on one cell: to any pair of adjacent points $\{n,m+1\} $,
$\{n+1,m\} $ on a diagonal we associate the polynomial
\begin{eqnarray}
\begin{cases}
\label{biqua}
 \kappa_{n,m}& =  ( a_1 a_7 - a_3^2 + a_4^2 - a_5^2) \; X_{n,m+1}
 Y_{n,m+1} X_{n+1,m} Y_{n+1,m} \\ & + ( a_1 a_6 + a_2 a_4 - a_2 a_3 -
 a_2 a_5 )\; ( X_{n,m+1}^2 Y_{n+1,m} X_{n+1,m} + X_{n+1,m}^2 X_{n,m+1}
 Y_{n,m+1}) \\ & +( a_2 a_7 + a_4 a_6 - a_3 a_6 - a_5 a_6 )\; (
 Y_{n,m+1}^2 Y_{n+1,m} X_{n+1,m} + Y_{n+1,m}^2 X_{n,m+1} Y_{n,m+1})
 \\ & + ( a_2 a_6 - a_3 a_5) \;( X_{n,m+1} ^2 Y_{n+1,m}^2 +
 Y_{n,m+1}^2 X_{n+1,m}^2) \\ & + ( a_1 a_4 - a_2^2)\; X_{n,m+1}^2
 X_{n+1,m}^2 + ( a_4 a_7 - a_6^2 ) \; Y_{n,m+1}^2 Y_{n+1,m}^2
\end{cases}
\end{eqnarray}
The key fact is that these polynomials split into two factors as soon
as $n+m\geq 2$. One of these two factors is common to $X_{n+1,m+1}$
and $Y_{n+1,m+1}$ the second one is common to $X_{n+2,m+2}$ and
$Y_{n+2,m+2}$. The evolution equations may be rewritten as the
recurrence
\begin{eqnarray}
\boxed{
\begin{split}
%\begin{cases}
\label{hiroqV}
 \Omega_{n,m}{\bf  \Omega_{n+1,m+1}}  &= \kappa_{n-1,m-1} \\ 
{\bf  X_{n+1,m+1} \; \Omega_{n+1,m+1}}  &=  - a_{{2}}X_{{n,m+1}}X_{{n,m}}X_{{n+1,m}}
- a_{{3}} X_{{n,m}}X_{{n+1,m}}Y_{{n,m+1}} \\
& - a_{{4}} X_{{n,m+1}}X_{{n+1,m}}Y_{{n,m}} - a_{{5}} X_{{n,m+1}} X_{{n,m}}Y_{{n+1,m}}
 -a_{{7}}Y_{{n,m+1}}Y_{{n,m}}Y_{{n+1,m}} \\ &
-  a_{{6}} \left( X_{{n,m+1}}Y_{{n,m}}Y_{{n+1,m}}+X_{{n,m}}Y_{{
n,m+1}}Y_{{n+1,m}}+X_{{n+1,m}}Y_{{n,m+1}}Y_{{n,m}} \right) \\
{\bf Y_{n+1,m+1} \; \Omega_{n+1,m+1} }&= a_{{1}} X_{{n,m+1}}X_{{n,m}}X_{{n+1,m}} 
 + a_{{3}}X_{{n,m+1}}Y_{{n,m}}Y_{{n+1,m}} \\ 
& + a_{{4}}X_{{n,m}}Y_{{n,m+1}}Y_{{n+1,m}}
 + a_{{5}}X_{{n+1,m}}Y_{{n,m+1}}Y_{{n,m}} + a_{{6}}Y_{{n,m+1}}Y_{{n,m}}Y_{{n+1,m}} 
 \\
&
+ a_{{2}} \left( X_{{n,m+1}}X_{{n,m}}Y_{{n+1,m}} 
+ X_{{n,m}}X_{{n+1,m}}Y_{{n,m+1}} 
+ X_{{n,m+1}}X_{{n+1,m}}Y_{{n,m}} \right),
%\end{cases}
\end{split}
}
\end{eqnarray}
with the added initial condition that $\Omega_{k,l}=1$ for
$k+l=1$. The various points entering the defining relation of the
$\Omega$'s are pictured in Figure 1 with the curved arrows.  The
$\Omega$'s and the $X,Y$'s given by (\ref{hiroqV}) are {\em
  polynomials in the initial conditions, and there is no additional
  common factor to $X_{n+1,m+1}$ and $Y_{n+1,m+1}$ for generic values
  of the parameters $a$}. Relations (\ref{hiroqV}) yield the sequence
of degrees of the iterates found in~\cite{Vi06,Vi08}, quadratic growth
and vanishing entropy.  More details will be given elsewhere.

%%%%%%%%%%%%%%%%%%%%%%%%%%%%%%%%%%%%%%%%%%%%%%%%%%%%%%%%%%%%%%%%%%%%%%%%%%%%

\section{Conclusion and perspectives}
 
We have shown that, for systems undergoing a rational discrete
evolution, a self-organisation takes place after a finite number of
steps: the structure of the iterates stabilises - this is to be
compared with the results of~\cite{CrJo02} - and the $\tau$ functions
pop out spontaneously as pieces of the components of the iterates,
providing a new set of variables to describe the evolution.

This change of description of the models is not to be confused with a
usual (birational) change of coordinates: we barter the original
coordinates for pieces of the components of { strings of successive
  iterates and transforms of the factors of the multipliers
  $\kappa_\varphi$}.

The recurrence relations obeyed by the new variables provide us with
an exact evaluation of the algebraic entropy and of its avatars
obtained by reductions to integers and finite fields, without
restriction to integrability. They support, by the form they take, the
fundamental conjecture presented in~\cite{BeVi99} that the entropy is
always the logarithm of an algebraic integer. They also invite us to
make contact with the results of~\cite{ChIsAs98,AdMo99} on orthogonal
polynomials.

In the integrable cases our approach may finally yield, in addition to
the known applications (Lax pairs, special solutions, soliton
solutions, grassmanian description) a new classification tool.

All of this is matter for future work.

\bigskip

{\bf Acknowledgements}. I would like to thank J. Hietarinta and
DJ. Zhang for fruitful exchanges during the elaboration of this
work. I would like to thank N. Joshi for stimulating discussions,
hospitality and support at the occasion of the Second Integrable
Systems Workshop, School of Mathematics and Statistics University of
Sydney, December 2014, where a part of these results was presented.

%\bibliography{../../ref}

\begin{thebibliography}{10}

\bibitem{SIDE}
Side: Symmetries and integrability of difference equations.
\newblock http://www.side-conferences.net/.

\bibitem{Ko89}
S.V. Kovalevska, {\em Sur le probl\`eme de la rotation d'un corps solide autour
  d'un point fixe}.
\newblock Acta Math. {\bf 12} (1889), pp. 177--232.

\bibitem{Pa02}
P.~Painlev\'e, {\em Sur les \'equations diff\'erentielles du second ordre et
  d'ordre sup\'erieur dont l'int\'egrale est uniforme}.
\newblock Acta Mathematica  (1902), pp. 1--85.

\bibitem{Ok79}
K.~Okamoto, {\em Sur les feuilletages associ\'es aux \'equations du second
  ordre \`a points critiques fixes de P. Painlev\'e}.
\newblock Jap. J. Math  (1979), pp. 1--79.

\bibitem{HiKr92}
J.~Hietarinta and M.~Kruskal.
\newblock Hirota forms for the six painlev\'e equations from singularity
  analysis.
\newblock In D.~Levi and P.~Winternitz, editors, {\em Painlev\'e
  Transcendents:Their Asymptotics and Physical Applications (NATO ASI B278)},
  pages 175--185, New York,  (1992). Plenum Press.

\bibitem{GrRaPa91}
B.~Grammaticos, A.~Ramani, and V.~Papageorgiou, {\em Do integrable mappings
  have the {P}ainlev\'e property?}
\newblock Phys.\ Rev.\ Lett. {\bf 67} (1991), pp. 1825--1827.

\bibitem{Sa01}
H.~Sakai, {\em Rational Surfaces Associated with Affine Root Systems and
  Geometry of the Painlev\'e Equations}.
\newblock Comm. Math. Phys. {\bf 220}(1) (2001), pp. 165--229.

\bibitem{QuRoTh88}
G.R.W. Quispel, J.A.G. Roberts, and C.J. Thompson, {\em Integrable Mappings and
  Soliton Equations}.
\newblock Phys.\ Lett. {\bf A 126} (1988), p. 419.

\bibitem{QuRoTh89}
G.R.W. Quispel, J.A.G. Roberts, and C.J. Thompson, {\em Integrable Mappings and
  Soliton Equations II}.
\newblock Physica {\bf D34} (1989), pp. 183--192.

\bibitem{Du10}
J.J. Duistermaat.
\newblock {\em Discrete Integrable Systems}.
\newblock Monographs in Mathematics. Springer,  (2010).

\bibitem{Ar90}
V.I. Arnold, {\em Dynamics of complexity of intersections}.
\newblock Bol. Soc. Bras. Mat. {\bf 21} (1990), pp. 1--10.

\bibitem{BeVi99}
M.~Bellon and C-M. Viallet, {\em Algebraic Entropy}.
\newblock Comm. Math. Phys. {\bf 204} (1999), pp. 425--437.
\newblock chao-dyn/9805006.

\bibitem{Vi08b}
C-M.Viallet, {\em Algebraic dynamics and algebraic entropy}.
\newblock International Journal of Geometric Methods in Modern Physics {\bf
  5}(8) (2008), pp. 1373--1391.

\bibitem{Ha05}
R.~Halburd, {\em Diophantine integrability}.
\newblock J. Phys. A {\bf 38}(16) (2005), pp. L263--L269.
\newblock arXiv:nlin.SI/0504027.

\bibitem{AnMaVi05}
J-C.~Angl\`es d'Auriac, J-M. Maillard, and C-M Viallet, {\em On the complexity
  of some birational transformations}.
\newblock J.Phys. A {\bf 39} (2006), pp. 3641--3654.
\newblock arXiv:math-ph/0503074.

\bibitem{RoVi03}
J.A.G. Roberts and F.~Vivaldi, {\em Arithmetical method to detect integrability
  in maps}.
\newblock Phys. Rev. Lett. {\bf 90} (2003), pp. 034102--1--034102--4.

\bibitem{RoJoVi03}
J.A.G. Roberts, D.~Jogia, and F.~Vivaldi, {\em The Hasse-Weil bound and
  integrability detection in rational maps}.
\newblock J. of Nonlinear Math. Phys. {\bf 10}(Supplement 2) (2003), pp.
  166--180.

\bibitem{Si07}
J.H. Silverman.
\newblock {\em The Arithmetic of Dynamical Systems}.
\newblock Number 241 in Graduate Texts in Mathematics. Springer-Verlag,
  (2007).

\bibitem{Mc07}
C.T. McMullen, {\em Dynamics on blowups of the projective plane}.
\newblock Publ. Math. Inst. Hautes Etudes Sci. {\bf 105} (2007), pp. 49--89.

\bibitem{DiHaKaKo14}
G.~Dimitrov, F.~Haiden, L.~Katzarkov, and M.~Kontsevich.
\newblock {\em Dynamical systems and categories}.
\newblock arXiv:1307.841,  (2013).

\bibitem{Hi77}
R.~Hirota, {\em Nonlinear partial difference equations.I,II,III}.
\newblock Journal of the Physical Society of Japan {\bf 43} (1977), pp.
  1424,2074,2079.

\bibitem{Sato81}
M.~Sato, {\em Soliton equations as dynamical systems on infinite-dimensional
  Grassmann manifold}.
\newblock RIMS-Kokyuroku {\bf 439} (1981), pp. 30--46.

\bibitem{SaSa82}
M.~Sato and Y.~Sato, {\em Soliton equations as dynamical systems on infinite
  dimensional {G}rassmann manifold}.
\newblock Lecture Notes Appl. Anal. {\bf 5} (1982), pp. 259--271.

\bibitem{Hi81}
R.~Hirota, {\em Discrete Analogue of a Generalized Toda Equation}.
\newblock J. Phys. Soc. Japan {\bf 50} (1981), pp. 3781--3791.

\bibitem{Mi82}
T.~Miwa, {\em On Hirota's difference equations}.
\newblock Proc. Japan Acad. Ser. A Math. Sci. {\bf 58}(1) (1982), pp. 9--12.

\bibitem{DaJiMi00}
E.~Date, M.~Jimbo, and T.~Miwa.
\newblock {\em Solitons: Differential equations, symmetries and infinite
  dimensional algebras}.
\newblock Cambridge Tracts in Mathematics 135. Cambridge University Press,
  (2000).

\bibitem{GrRaHi94}
B.~Grammaticos, A.~Ramani, and J.~Hietarinta, {\em Multinear operators: the
  natural extension of Hirota's bilinear formalism}.
\newblock Phys. Lett. {\bf A}(190) (1994), pp. 65 -- 70.

\bibitem{FoZe02}
S.~Fomin and A.~Zelevinsky, {\em The Laurent Phenomenon}.
\newblock Advances in Applied Mathematics {\bf 28} (2002), pp. 119,144.

\bibitem{HaPr05}
B.~Hasselblatt and J.~Propp, {\em Degree-growth of monomial maps}.
\newblock Ergodic Theory and Dynamical Systems {\bf 27}(05) (2007), pp.
  1375--1397.
\newblock arXiv:math.DS/0604521.

\bibitem{Ho07}
A.N.W. Hone, {\em Laurent polynomials and superintegrable maps.}
\newblock SIGMA Symmetry Integrability Geom. Methods Appl. {\bf 3} (2007).

\bibitem{FoHo11}
A.P. Fordy and A.~Hone, {\em Symplectic maps from cluster algebras}.
\newblock Sigma {\bf 7} (2011), p. 091.

\bibitem{Fo14}
A.~Fordy.
\newblock {\em Periodic cluster mutations and related integrable maps}.
\newblock arXiv:1403.8061.

\bibitem{Prxx}
J.~Propp.
\newblock The somos sequence site.
\newblock http://faculty.uml.edu/jpropp/somos.html.

\bibitem{FaVi93}
G.~Falqui and C.-M. Viallet, {\em Singularity, complexity, and
  quasi--integrability of rational mappings}.
\newblock Comm.\ Math.\ Phys. {\bf 154} (1993), pp. 111--125.
\newblock hep-th/9212105.

\bibitem{Mc71}
E.M. McMillan.
\newblock A problem in the stability of periodic systems.
\newblock In E.~Britton and H.~Odabasi, editors, {\em A tribute to E.U.
  Condon}, Topics in Modern Physics, pages 219--244, Boulder,  (1971). Colorado
  Assoc. Univ. Press.

\bibitem{RaGrSa95}
A.~Ramani, B.~Grammaticos, and J.~Satsuma, {\em Bilinear discrete Painlev\'e
  equations}.
\newblock J. Phys. A: Math. Gen. {\bf 28} (1995), pp. 4655--4665.

\bibitem{OhRaGrTa96}
Y.~Ohta, A.~Ramani, B.~Grammaticos, and K.M. Tanizhmani, {\em From discrete to
  continuous Painlev\'e equations}.
\newblock Phys. Lett. {\bf A}(216) (1996), pp. 255--261.

\bibitem{KaMaOi00}
K.~Kenjiwara, K.~Maruno, and M.~Oikawa, {\em Bilinearisation of discrete
  soliton equations through the singularity confinement test}.
\newblock Chaos, Solitons and Fractals {\bf 11} (2000), pp. 33--39.

\bibitem{SaKaGrHiRa95}
J.~Satsuma, K.~Kajiwara, B.~Grammaticos, J.~Hietarinta, and A.~Ramani, {\em
  Bilinear discrete Painlev\'e and its particular solutions}.
\newblock J. Phys A: Math. Gen. {\bf 28} (1995), pp. 3541--3548.

\bibitem{JiSa96}
M.~Jimbo and H.~Sakai, {\em A $q$-analog of the sixth Painlev\'e equation}.
\newblock Letters in Mathematical Physics {\bf 38} (1996), pp. 145--154.

\bibitem{JaMaNo98}
F.~Jaeger, M.~Matsumoto, and K.~Nomura, {\em Bose-Mesner algebras related to
  type II matrices and spin models}.
\newblock Journal of Algebraic Combinatorics {\bf 8} (1998), pp. 39--72.

\bibitem{HiVi98}
J.~Hietarinta and C.-M. Viallet, {\em Singularity confinement and chaos in
  discrete systems}.
\newblock Phys. Rev. Lett. {\bf 81}(2) (1998), pp. 325--328.
\newblock solv-int/9711014.

\bibitem{Ta01c}
T.~Takenawa, {\em A geometric approach to singularity confinement and algebraic
  entropy}.
\newblock J. Phys. A: Math. Gen. {\bf 34} (2001), pp. L95--L102.

\bibitem{PePfSu11}
M.~Petrera, A.~Pfadler, and Y.B. Suris, {\em On integrability of Hirota-Kimura
  type discretizations}.
\newblock Regular and Chaotic Dynamics {\bf 16}(3-4) (2011), pp. 245--289.
\newblock arXiv:1008.1040.

\bibitem{AbHaHe00}
M.J. Ablowitz, R.~Halburd, and B.~Herbst, {\em On the extension of the
  Painlev\'e property to difference equations}.
\newblock Nonlinearity {\bf 13} (2000), pp. 889--905.

\bibitem{QuCaSa92}
G.R.W. Quispel, H.W. Capel, and R.~Sahadevan, {\em Continous symmetries of
  differential-difference equations: the Kac-van Moerbeke equation and the
  Painlev\'e reduction}.
\newblock Phys. Lett. {\bf A}(170) (1992), pp. 379--383.

\bibitem{Ma74}
S.V. Manakov, {\em Complete integrability and stochastization of discrete
  dynamical systems}.
\newblock Zh. Eksp. Teor Fiz {\bf 67} (1974), pp. 543--555.
\newblock Sov. Phys. JETP, Vol. 40, No 2, pp 269--274.

\bibitem{Vi14}
C.M. Viallet.
\newblock {\em Algebraic entropy for differential-delay equations}.
\newblock arXiv:1408.6161.

\bibitem{Vi06}
C-M. Viallet.
\newblock {\em Algebraic entropy for lattice equations}.
\newblock arXiv:math-ph/0609043.

\bibitem{Vi08}
C-M. Viallet, {\em Integrable lattice maps: $Q_V$, a rational version of
  $Q_4$}.
\newblock Glasgow Math. J. {\bf 51 A} (2009), pp. 157--163.
\newblock arXiv:0802.0294.

\bibitem{AdBoSu03}
V.E. Adler, A.I. Bobenko, and Yu.B. Suris, {\em Classification of integrable
  equations on quad-graphs. The consistency approach}.
\newblock Comm. Math. Phys. {\bf 233}(3) (2003), pp. 513--543.
\newblock arXiv:nlin.SI/0202024.

\bibitem{AdBoSu04}
V.E. Adler, A.I. Bobenko, and Yu.B. Suris, {\em Geometry of Yang-Baxter maps:
  pencils of conics and quadrirational mappings}.
\newblock Commun. Anal. Geom. {\bf 12} (2004), pp. 967--1007.
\newblock arXiv:math.QA/0307009.

\bibitem{AdBoSu07}
V.E. Adler, A.I. Bobenko, and Yu.B. Suris, {\em Discrete nonlinear hyperbolic
  equations. Classification of integrable cases}.
\newblock Funct. Anal. Appl. {\bf 43} (2009), pp. 3--17.
\newblock arXiv:0705.1663.

\bibitem{MiWaXe11}
A.~V. Mikhailov, J.~P. Wang, and P.~Xenitidis, {\em Recursion operators,
  conservation laws, and integrability conditions for difference equations}.
\newblock Teoret. Mat. Fiz. {\bf 1} (2011), pp. 23--49.
\newblock transl. Theoret. and Math. Phys. 167 (2011), pp 421--443,
  arXiv:1004.5346.

\bibitem{MiWa11}
A.~V. Mikhailov and J.~P. Wang, {\em A new recursion operator for Adlerʼs
  equation in the Viallet form}.
\newblock Physics Letters {\bf A 375} (2011), pp. 3960--3963.
\newblock arXiv:1105.1269.

\bibitem{CrJo02}
C.~Cresswell and N.~Joshi, {\em Consistent composition of B\"acklund
  transformations produces confined maps}.
\newblock Lett. Math. Phys. {\bf 61} (2002), pp. 1--14.

\bibitem{ChIsAs98}
Y.~Chen, M.E.H. Ismail, and W.~van Assche, {\em Tau-function construction of
  the recurrence coefficients of orthogonal polynomials}.
\newblock Adv. Applied Math. {\bf 20} (1998), pp. 141--168.

\bibitem{AdMo99}
M.~Adler and P.~van Moerbeke, {\em Generalized orthogonal polynomials, discrete
  KP and Riemann-Hilbert problems}.
\newblock Comm. Math. Phys. {\bf 207} (1999), pp. 589--620.
\newblock arXiv:nlin/0009002.

\end{thebibliography}

\end{document}